\documentstyle[preprint,aps,prd]{revtex}
\def\be{\begin{equation}}
\def\ee{\end{equation}}
\def\bea{\begin{eqnarray}}
\def\eea{\end{eqnarray}}
\def\l{\label}
\def\c{\cite}
\def\r{\ref}
\def\th{\theta}

\begin{document} 
\draft 
\tighten

\preprint{\vbox{\hbox{SUSX-TH-96-006}}}

\title{Bubble collisions in Abelian gauge theories and the geodesic rule.}

\author{E.J. Copeland\thanks{E=mail: E.J.Copeland@sussex.ac.uk} \& 
        P.M. Saffin\thanks{E-mail: p.m.saffin@sussex.ac.uk}} 
\address{School of Mathematical and Physical Sciences, University of
Sussex, \\Brighton BN1 9QH, United Kingdom}

\date{\today} \maketitle

\begin{abstract} 
In an Abelian gauge symmetry, spontaneously broken at a first 
order 
phase transition, we investigate the evolution of two and three 
bubbles of the broken 
symmetry phase. The full field equations are evolved and we concentrate in 
particular on 
gauge invariant quantities, such as the magnetic field
and the integral around a closed loop
of the phase gradient.  
An intriguing feature emerges, namely, the 
geodesic rule, commonly used in numerical simulations to determine the 
density 
of defects formed is shown not to hold in a number of circumstances. It 
appears 
to be a function of the initial separation of the bubbles, and the 
coupling 
strength of the gauge field. The reason for the breakdown 
is that in the collision region the radial mode {\it can} be excited and {\it 
often} oscillates about its symmetry  restoring value rather than settling 
to 
its broken symmetry value. This can lead to extra windings being 
induced in these regions, hence extra defects (anti-defects) being formed. 
\end{abstract}

\pacs{98.80.Cq}

\narrowtext

\noindent There is much evidence that the early Universe was characterised 
by a 
series of phase transitions in which a high energy `old' symmetry phase 
was 
spontaneously broken to a low energy `new' symmetry phase, possibly 
leading to the formation of topological defects. 
Such objects are also readily found in condensed 
matter 
systems (although of a much lower energy scale) \cite{nato94}, 
so although direct observational evidence for them 
in 
cosmology is still in doubt there exists plenty of evidence for their 
existence 
in terrestrial experiments. Given they could exist, we need to know their 
initial 
distribution so that we can determine the cosmological implications of the 
defects.

In this article we are concerned with the dynamics of Abelian-Higgs 
fields 
in a first order phase transition when bubbles of the new phase are 
nucleated in 
the background of the old phase. Topological defects form in the regions 
where 
the bubbles collide, so it would be useful to investigate the behaviour of 
the 
fields in this region. We will be following the work of 
\cite{hawking82} in determining the evolution of the bubble walls, but 
will be  concerning ourselves with defect formation 
rather than the rate at which the energy in the bubble walls 
would be thermalized. 

We will be considering a $U(1)$ theory with a complex scalar field $\Phi$, 
and Lagrangian
\be
{\cal L} = \left(D_\mu \Phi\right)^* D^\mu \Phi - \frac14 F_{\mu \nu}F^{\mu \nu} 
- 
V(\Phi)
\l{lag}
\ee
where $D_\mu \Phi = \partial_\mu \Phi - ie A_\mu \Phi,\,F_{\mu \nu} = 
\partial_\mu A_\nu - \partial_\nu A_\mu$, $e$ is the gauge coupling 
constant and 
the potential $V$ is a first order potential, a function of $| \Phi |^{2}$ 
with a 
local minimum at $| \Phi | = 0$ and a global minimum at $| \Phi | = 
\eta/\sqrt{2}$. The 
dynamics of the phase transition proceed as follows. As the Universe 
expands and 
cools, tunnelling can occur from the old 
phase $ 
\Phi = 0$ to the new phase where $\Phi \simeq \frac{\eta}{\sqrt{2}} e^{i\th}$. 
The 
symmetry has been spontaneously broken and within each bubble there is a 
random 
choice of the phase angle $\th$. The bubbles, nucleated at random 
points expand and collide, with their nucleation rate being determined 
from the bounce solution to the Euclidean action\c{coleman77}. 

The first 
definitive explanation for defect formation 
was provided in the work of Kibble 
\cite{kibble76}. Consider the case of cosmic strings. 
It is assumed that within each bubble the phase $\th$ is 
constant, with neighbouring bubbles being uncorrelated. When 
two 
bubbles with phases $\th_1$ and $\th_2$ meet, any discontinuity between the 
phases is smoothed out. On energetic 
grounds, the shorter path between $\th_1$ and $\th_2$ on the vacuum
manifold is chosen, a result 
known as the `geodesic rule'. For the collision of three bubbles, a string 
may be trapped in the region between the bubbles. This depends on 
the net 
phase change in going sequentially $\th_1 \rightarrow \th_2 \rightarrow 
\th_3 \rightarrow \th_1$ (see fig 8). 

In a series of papers \cite{sriv92}, 
Srivistava demonstrated that the geodesic rule does indeed hold for global 
theories. Recently though, with Rudaz \cite{rudaz93}, he questioned the 
reliability of the rule when gauge fields are present, pointing out that it may not 
make sense to talk about phase differences 
between bubbles in a gauge theory. It is possible to gauge transform the 
phase 
difference to any value. As a consequence they argued that string 
formation in such gauge theories may well be strongly suppressed as 
compared to 
the global theory case. 
Hindmarsh et al \c{hindmarsh94} analysed the bubble collisions using an 
analytic 
approach and concluded that the geodesic rule did actually hold. They 
pointed out that the geodesic rule emerges not from energy considerations 
but 
from the equations of motion themselves-- the dynamics. Recently Kibble and 
Vilenkin \c{kibble95}, also using an analytic approach addressed the same 
issue, 
concluding that the geodesic rule nearly always held. They went further, 
including dissipation terms induced by the finite plasma conductivity that 
the 
bubbles expand in, they demonstrated that this can cause the phases to 
equilibrate on a timescale much smaller than the bubble radii at the 
time of 
collision. Such a result seems to vindicate the common assumption 
that the geodesic rule holds. However, as the authors stressed in 
\c{hindmarsh94} and \c{kibble95}, throughout the calculations various 
assumptions have to be made about the behaviour of the fields. 
For example in \c{kibble95} it was assumed that the 
radial 
mode of the Higgs field is strongly damped, settling into its equilibrium 
value 
on a timescale short compared to the phase equilibrium process. 
A similar assumption was made in \c{hindmarsh94} where the 
variation of the radial mode was dropped inside the bubble.
In general 
though, this may not happen.In this article 
we 
report on a project in which we solved the full field evolution equations 
for 
two colliding bubbles numerically, keeping track on the variation of the 
radial 
as well as the phase degrees of freedom. The results
open up 
the possibility that the geodesic rule may not be as widespread as first 
thought. 

Following \c{hawking82} we consider a potential which is a function of 
$|\Phi|^2$ and which has a local minimum at the origin and a circle of 
degenerate global minima away from the origin. The simplest such potential 
is a cubic in $|\Phi|^2$, hence we consider 
\be
V(\phi)=a(|\phi|^{2}+B)(|\phi|^{2}-c^{2})^{2}.
\l{pot1}
\ee
Two of the three parameters ($a$, $c$) in Eqn.~(\ref{pot1}) can be set to 
unity by redefining the fields and coordinates.

The nucleation of a single bubble follows the bounce solution 
\c{coleman77}.The gauge fields are taken to vanish and the phase is required 
to be constant within the bubble.
The bounce solution found numerically was seen to compare favourably with a $\tanh(x)$ profile,
which was subsequently used in the simulations as the initial condition on the 
Higgs field. (Our results are robust to this approximation 
due to the Lorentz contraction of the bubble walls.)
In the problem of two 
bubbles colliding the tunnelling solution has $SO(2,1)$ symmetry
and a suitable coordinate system to utilise this is

\begin{eqnarray}
t&=&s \cosh(\psi)\\
y&=&s \sinh(\psi) \sin(\varphi)\\
z&=&s \sinh(\psi) \cos(\varphi)\\
x&=&x
\end{eqnarray}

The fields must then evolve independantly of the trigonometric and
hyperbolic angles $\varphi$ and $\psi$. 
This greatly simplifies the calculation since the solution becomes a 
function of two variables, the x coordinate, and the `time' $s$ where 
$s^{2}=t^{2}-y^{2}-z^{2}$. A stationary point, $s,~x=$ const, in these 
coordinates thus corresponds to an expanding loop in Minkowski space, 
with the $x$ axis as the symmetry axis.

There are two useful representations for $\Phi$. Firstly one has
$\Phi=\frac{\rho}{\sqrt{2}}e^{i\theta}$. This is easier to visualise
but difficult to implement numerically, due to the ambiguity in $\theta$ when
the field modulus $\rho$ vanishes. Secondly there is a Cartesian description
$\Phi=\frac{1}{\sqrt{2}}(u+iv)$, where $u$, $v$ are real.
The choice of the Lorentz gauge ($\nabla_{\alpha}A^{\alpha}=0$) and the 
independance of $\psi$, $\varphi$ leads to the set of evolution equations. 
\begin{eqnarray}
\l{rhoeqn}                                      
\ddot{\rho}&=&\rho''-\frac{2}{s}\dot{\rho}
             +\rho[\dot{\theta}^{2}-\theta'^{2}]
             -\frac{\partial {\cal V}}{\partial \rho}
            +e^{2}\rho[A_{s}^{2}-A_{x}^{2}]  
            +2e\rho[A_{x}\theta'-A_{s}\dot{\theta}]\\
\l{thetaeqn}  
\ddot{\theta}&=&\theta''-\frac{2}{s}\dot{\theta}
              -\frac{2}{\rho}[\dot{\rho}\dot{\theta}-\rho'\theta']   
              +\frac{2e}{\rho}[A_{s}\dot{\rho}-A_{x}\rho']\\
\l{aseqn}
\ddot{A_{s}}&=&A_{s}''-\frac{2}{s}\dot{A_{s}}+\frac{2}{s^{2}}A_{s}
             -e^{2}\rho^{2}A_{s}+e\rho^{2}\dot{\theta}\\
\l{axeqn}
\ddot{A_{x}}&=&A_{x}''- \frac{2}{s}\dot{A_{x}} -e^{2}\rho^{2}A_{x}
             +e\rho^{2}\theta'
\end{eqnarray} 
Where $\dot{f}=\partial f/\partial s$, $f'=\partial f/\partial x$.
Note that there is a term in ({\ref{thetaeqn}) depending
on $e$, the gauge fields and the derivatives of $\rho$, which may be 
interpreted as a forcing term for $\theta$ and is 
the central difference between taking $\rho$ as constant \c{kibble95} and
considering the global case \c{sriv92}.

For the initial conditions of two bubbles with a phase difference, the 
description of the field's modulus is easy to understand. As the bubbles
collide the intersection forms a ring, described by $s$, $x=$ constant, 
and the large amount of energy 
available in the walls at collision allows the $U(1)$ symmetry to be
restored in this ring. It is this localised symmetry restoration which
allows (but does not require) the phase to have a non-trivial winding 
about the ring, thus indicating the break down of the geodesic rule.  
Such a winding occurs in many cases. It is found that the amount of 
winding and its direction depends on the size of 
the gauge 
coupling, the initial separation and initial phase difference of the 
bubbles.
By holding the phase and bubble separation constant we may investigate
the $e$ dependance. When $e=0$, the global case, the
winding around the ring is seen to vanish, consistent with the geodesic rule. 
Increasing $e$ leads to a generation of winding, which can flip sign from being
positive to negative and also produce multiple $2\pi$ windings. Subsequent 
flips of sign occur for values of $e$ which are integer multiples of 
the value of $e$ for which the flip first occurs.

The dependance on initial separation is probed by holding $e$ and the
phase difference
constant. It is found that the greater the
separation,
the greater the winding. As for the dependance on the initial phase separation,
we observe that there is a cutoff around $\frac{\pi}{2}$ above which no 
winding is generated.

The picture describing the evolution of the Higgs field 
is shown for example in Figure 1. Two 
regions of true vacuum expand, collide and form a pocket of restored $U(1)$ 
symmetry which subsequently collapses and forms another pocket. These 
oscillations continue until the energy has been radiated away \c{hawking82}.

A non-trivial winding is observed to occur over the time period that the 
first pocket
collapses (see Figure 2).
This winding occurs at constant $s$, $x$ inside the true vacuum region
and so corresponds to an expanding loop following the intersection of the 
two bubbles. This loop has a unit winding and so corresponds to an expanding
loop of string. 
This timescale can be estimated \c{hawking82} 
by considering the thin wall approximation for a global theory.   
Let $s_{0}$ be the coordinate time at which the walls
collide and
$s_{2}$ when the pocket collapses. By matching the wall velocities before
and after collision (consistent with simulations) then it is possible to show:
\be
s_{2} \simeq \left[2\left(2\left[1-\frac{1}{3} \alpha^{2}\right]\right)
      ^{\frac{1}{3}}-1\right]s_{0}
\l{pocket}
\ee                                   
where $\alpha$ is the phase separation between the bubbles. 
The limiting case is when $s_{2}\simeq s_{0}$ for which we find a value of
$\alpha$ of $\simeq \sqrt{\frac{3}{2}} \simeq \frac{\pi}{2}$, a result that 
explains why
no windings are found above a certain initial phase separation. This phase
angle cutoff
is due to the extra energy in the phase wave that propogates away
from the collision region, leaving insufficient energy to restore
the symmetry. Shortly we 
shall see that the amount of winding depends on how long this first
pocket can survive. From Eqn.~({\r{pocket}) the lifetime varies
in proportion to $s_{0}$ so, since the bubbles reach 
relativistic velocities rapidly, the pocket's lifetime will 
vary in proportion
to the initial separation, a result that explains why more windings were found 
as the separation was increased.

Now we consider the cause of the winding. If we set the phase in the left hand 
bubble to be zero and in the right hand bubble 
to be
$0<\alpha<\pi$, the equations of motion require $A_{x}>0$ for all $x$ and 
$A_{s}<0$ for $(x>0)$ just after the collision. As the pocket initially 
expands
then $\dot{\rho}<0$, $\rho'>0$ which means the term
$\frac{2e}{\rho}\left(A_{s}\dot{\rho}-A_{x}\rho'\right)$
in Eqn.~(\ref{thetaeqn}) has components which oppose each other and lead to
a small forcing term for $\theta$. However, when the pocket starts to 
collapse then $\dot{\rho}>0$ and the components combine to create a 
negative forcing term for $\theta$, driving the field around the true vacuum
manifold and generating a winding (see Figures 2, 5).
Note that the longer the pocket survives, then 
the longer this forcing can act leading to more 
winding. An important feature of the winding
is that as the bubbles collide and the radial component of the field 
overshoots the global minima
restoring the symmetry, the field traces a twist in the
configuration space which leads to a winding of $+2\pi$ instead of the
$-2\pi$ which would occur without the twist (see Figures 4, 5 at s=31.5, s=32).

The sense of the winding depends upon the sign of the gauge fields.
If it is possible to change the sign of the gauge fields by the time the
pocket starts to collapse then the winding direction is reversed.
In equations (\ref{aseqn}, \ref{axeqn}) we find 
the harmonic term $-e^{2} \rho^{2}A$ which can generate 
a sign change on a timescale of $\frac{\pi}{e \rho}$. This scale is 
only significant for large $\rho$ and should be compared with the 
timescale of the modulus wall, $\delta$. If the gauge fields $A_{s},\,A_{x}$ can 
change sign in 
the time taken for the modulus to drop to zero, then they will retain 
their {\it new} sign until the pocket collapses. This leads to a change in the 
sense of the winding at $e \simeq \frac{\pi}{\eta \delta} n$, where $n$ 
is an integer. This demonstrates the sign flipping observed in the 
simulations at regular intervals of $e$ and agrees numerically with the 
observed periodicity. 

With a constant field modulus in the true vacuum, $\rho=\eta$, 
then the field equations
have the solution \mbox{$eA_{\mu}=\partial_{\mu}\theta$}.
Using the two bubble collision depicted in fig6, we may 
evaluate the associated magnetic flux of the winding. 
As the gauge field vanishes on sides BC, CD, DA then 
the magnetic flux, $\Phi=\oint \underline{A}.\underline{dl}$, inside 
the loop ABCD is $\Phi_{ABCD}=\int^{B}_{A}A_{x}$d$x$. 
When the collision region (the intersection of the two bubbles)
has moved from the x axis then along the $x$ axis 
$eA_{\mu} \simeq \partial_{\mu}\theta$
and so $\Phi \simeq \alpha_{12}/e+2 \pi n_{12}/e$. The value
of the integer $n_{12}$ represents the extra winding
and depends on how the geodesic rule is broken,
with the geodesic rule requiring $n_{12}$ to vanish.
From the simulations it was found that couplings of e=0.1, 0.5,
1.0, 1.5 (with an initial separation of 60 units and phase
difference of $\alpha_{12}=\pi /4$) lead to extra 
windings of 0, 1, 2, -1 respectively. The flux measured 
through the loop ABCD 
for these couplings is shown in fig7, in units of $2\pi/e$. 
The magnetic flux, which is a gauge invariant quantity,
through the enclosed region is seen to oscilate about
$2\pi/e(1/8+n_{12})$ with $n_{12}=0,1,2,-1$ thus confirming
the expected flux. The late time
oscillations correspond to the natural frequency 
$\omega=e\eta$.

It is important to know whether this breaking of the geodesic rule
rule can lead to stable defects.
As explained, the collision region forms a ring and this ring can
have a winding, i.e. a loop of string. This loop is close to the
false vacuum surrounding the bubbles but as seen in fig2 it 
cannot escape into the false vacuum, wiping out the string. In order
to form a string surrounded by true vacuum we need at least three
bubbles, we now consider how the above can alter the predictions
for these trapped strings.
For a path $\Gamma$ (fig8)
which is outside the collision regions then the field modulus assumes
a non zero value so that the phase is defined on each segment of
the path. The path $\Gamma$ also has the property that any leg of the 
path joining two bubbles has not been reached by the third bubble.
This means that we may treat the phase on each leg of $\Gamma$ simply
by considering the relevant two bubble collision.
The total winding enclosed in this path is then
$$n_{t}=\frac{1}{2\pi}\left(\alpha_{12}+\alpha_{23}+\alpha_{31}\right)
         +n_{12}+n_{23}+n_{31}$$
where $n_{ij}$ is the extra winding induced by the gauge field between
bubbles $i$ and $j$.
As an example, consider bubbles with $\theta_{1}=0$, $\theta_{2}=\pi/4$
and $\theta_{3}=\pi/2$, all initially separated by $60$ 
units with a gauge coupling
of $e=0.5$. From figs2,5 we know that as $\alpha_{12}=\alpha_{23}=\pi/4$, then
$n_{12}=n_{23}=1$. Also, because $\alpha_{31}=\pi/2$ then no extra winding 
is expected between bubbles 1, 3
due to the upper bound on $\alpha$ for extra winding caused by the
energy in the phase wave, so $n_{31}=0$. The total winding thus expected
is two, whereas the geodesic rule would have predicted no winding.
The result of a simulation is shown in fig9. The phase is seen to change
by $4\pi$ as we traverse the sides of the triangle and this leads to
the presence of the two single winding vortices which can be seen in
the figure. 
The existence of a path, $\Gamma$, where the field
is away from the collision area is important. If such a path cannot
be drawn then it may be possible for the extra windings to escape.

The effects of dissipation are expected to play a significant r\^ole
in this mechanism, as the field is required to have enough energy
locally to temporarily restore the symmetry. If this energy can be
dissipated away on short enough timescales then the geodesic rule
will be restored.

Following ref\cite{kibble95}, we modelled the effects of dissipation
using an Ohmic current, defined through the projection relation
$$\left[g^{\mu}_{\nu}-U_{\nu}U^{\mu}\right]j_{Ohm}^{\nu}
      =-\sigma U_{\nu} F^{\nu \mu}$$
where $U$ is the fluid four velocity.
We then use the conservation constraint 
\mbox{$\nabla_{\mu}j_{Ohm}^{\mu}=0$}
to fully define the current and rewrite the gauge field equations
of motion as 
\mbox{$\partial_{\mu}F^{\mu \nu}=j_{Noether}+j_{Ohm}$}.
The effects of so doing are seen in fig3 where, for the parameters,
used in figs2,5 the winding has
reverted back to the geodesic rule when the conductivity, $\sigma=0.5$. 
As the conductivity
of the plasma is increased then the more the gauge field gets coupled
to the Ohmic current rather than the Noether current, the effect
of this is
to reduce the gauge field forcing term of equation (\ref{thetaeqn})
and, for a large enough conductivity, the geodesic rule
is restored.
      
On the face of it the cosmological implications of a breakdown
of the geodesic rule could be very significant.  
Typically, as mentioned earlier, the phases are set at random in each bubble and 
the geodesic rule is used to determine the density of defects 
formed initially. This leads 
to a distribution of open and closed string. Now we seem to find that depending 
on the parameters, if the phase difference between two bubbles is less than 
about $\pi/2$, then it is likely that new windings will be induced in the 
collision region. A technique has recently been developed \cite{borrill96} in which the length 
distribution of strings can be calculated numerically, 
without a regular lattice dependance.
This method is particularly suited to modelling a first order 
phase transition and
it would be interesting to extend those results to take into account the
effects described above.

There is a precedent for extra vortices being found in the collision 
regions of the bubble walls. Digal and Srivastava \cite{sriv95} have analysed 
the behaviour of two bubbles colliding in a global U(1) theory (in 2+1 dimensions)
where the global 
symmetry is broken both spontaneously {\it and} explicitely. 
They find that in the 
coalesced region of the bubbles, field oscillations result in the 
production of a number of defects (vortices and anti-vortices).
We have been able to confirm this behaviour in the case of 3+1 dimensions
and will report on this elsewhere\cite{copeland96}.

In conclusion, the results we have obtained suggest that 
a set of rules exist for predicting 
the likely outcome of 
defects after two bubble collisions. 
By calculating gauge invariant quantities, like the magnetic flux
and closed loop integral of the phase gradient, we have shown how
the number of windings found varies.
The phase winding of the loop was seen to depend on the initial 
separation of the bubbles, the strength of the coupling 
constant $e$,  
and on the currently unknown strength of the dissipation term, $\sigma$.
A point to stress is that the rule determining winding will be
one that depends on the dynamics of the fields, 
not one determined
from energetic considerations, a comment also made in \cite{hindmarsh94}.
The loops that are generated in this process are extra vortices,
the final state of the phase transition is different to that
predicted with the geodesic rule. An important and as yet unresolved
issue concerns the statistical length distribution of string loops
in the network of defects.
Naively there would appear to be far more loops produced which do
not simply shrink away. As has been pointed out by Borrill \cite{borrill96}
such a result could have implications for the normalisation of the
string tension.
We have shown that these extra loops of string are expected when the
phase separation is less than $\simeq \pi/2$. This will be the
case on average for around half of the bubble pairings and may therefore
significantly change the length distribution of string.


\acknowledgments
We are grateful to J. Borrill, M. Hindmarsh, R. Holman, T.W.B. Kibble 
and A.M. Srivastava for useful comments and conversations. 
P.M.S. is grateful to G.Vincent for tutoring in C and to PPARC for 
financial support.  

\listoffigures

\noindent Figure 1. Field evolution plot for $e = 0.0,\,0<s<60,\,-30<x<30$.
The bubbles are nucleated at $(s,x)=(0,\pm 30)$ 
The magnitude of the Higgs field is represented by the arrow length and 
the phase by its direction.

\noindent Figure 2. Field evolution plot for $e = 0.5,\,0<s<60,\,-30<x<30$ 
where a winding of $2\pi$ is generated.

\noindent Figure 3. Field evolution plot for $e = 0.5, \sigma = 0.5 
\,0<s<60,\,-30<x<30$ 
where a winding of $2\pi$ is prevented due to the dissipation of the 
energy in the gauge field.

\noindent Figure 4. Sequence in field configuration $(u,v)$ space 
for $e = 0.0.$ and $\phi=1/\sqrt{2}(u+iv)$
along the path $-30<x<30$. 
   
\noindent Figure 5. Sequence in field configuration $(u,v)$ space 
for $e = 0.5.$ and $\phi=1/\sqrt{2}(u+iv)$
along the path $-30<x<30$. 

\noindent Figure 6. Schematic representation of a configuration for two
colliding bubbles with phases $\theta_{1}$, $\theta_{2} $
and phase difference $\alpha_{12}$
showing the path used to measure the magnetic flux.

\noindent Figure 7. Magnetic flux through the region depicted
in fig6 for various gauge couplings as a function of $s$. The
initial conditions are two bubbles nucleated at a separation
of $60$ units with a phase separation of $\pi/4$.

\noindent Figure 8. Schematic representation of a 
three bubble collision showing
bubbles with phases $\theta_{i}$ and phase separations
$\alpha_{ij}$.

\noindent Figure 9. Plot of a three bubble collision in the 
$x-y$ plane with
$e=0.5$ showing the generation of two stable windings which
contradicts the prediction of the geodesic rule. The three 
bubbles (with initial phases $0,\pi/4,\pi/2$) were nucleated 
at equal
times with their centres at the vertices of the equalateral
triangle. The figure represents the state which such a 
configuration will evolve to.

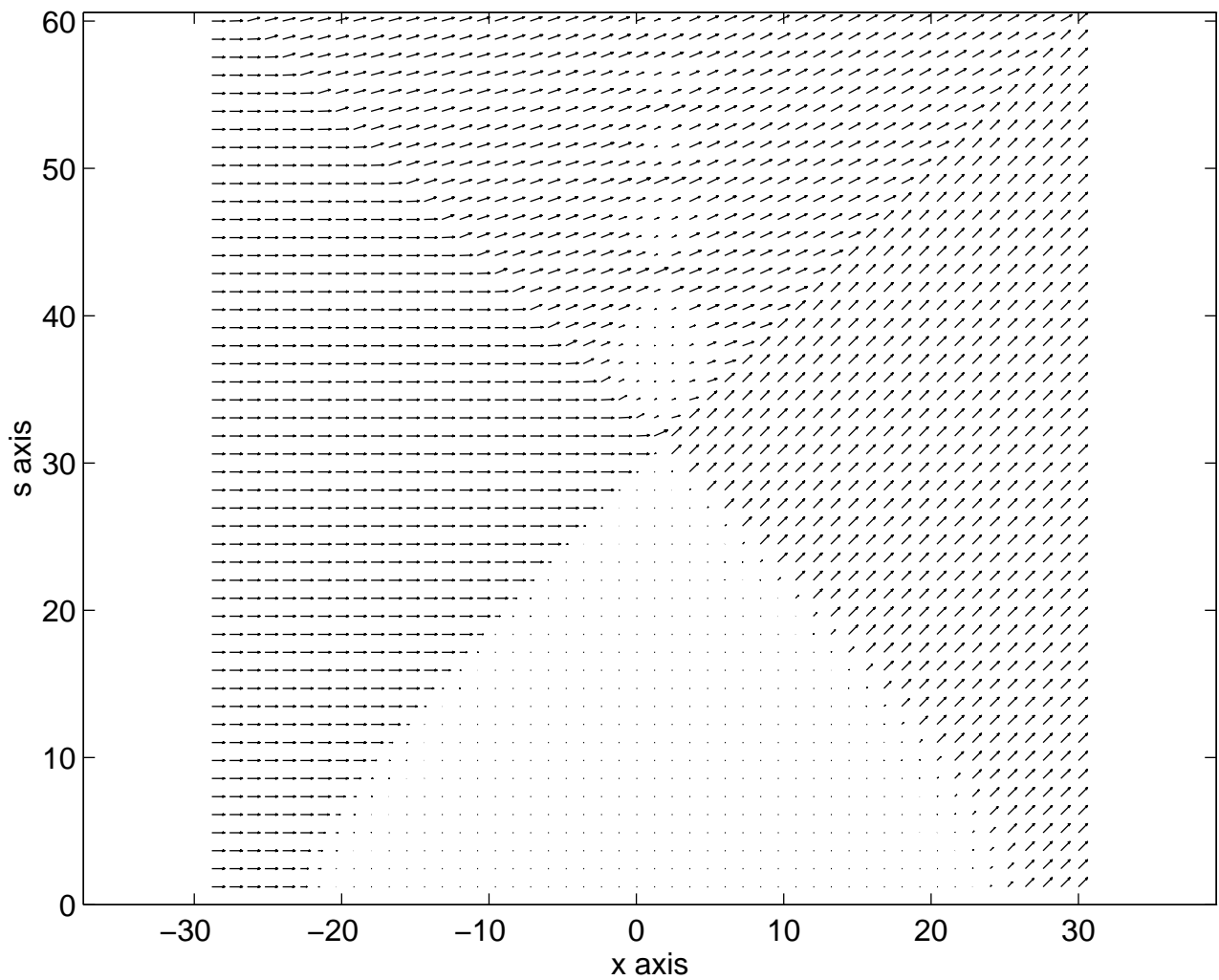
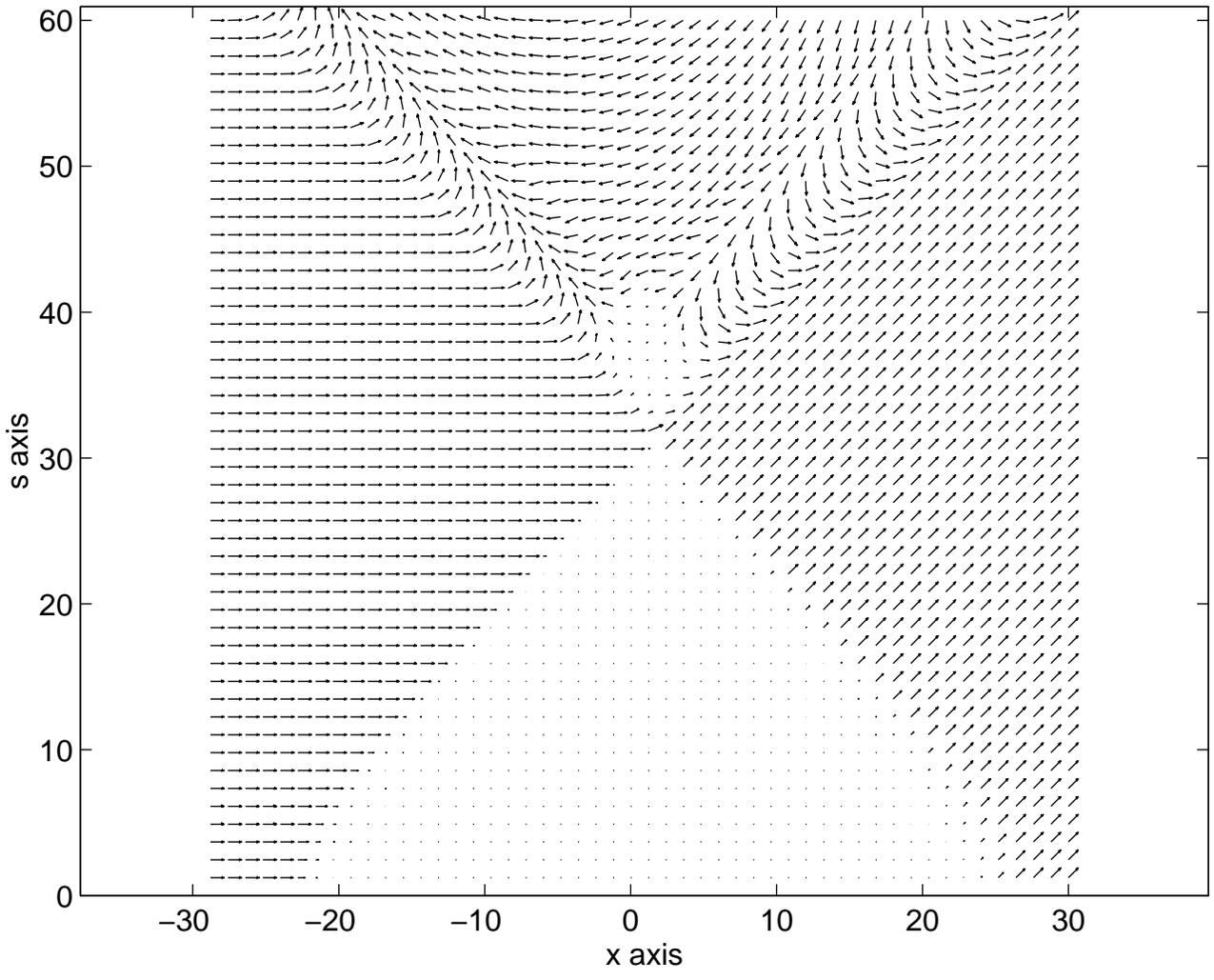
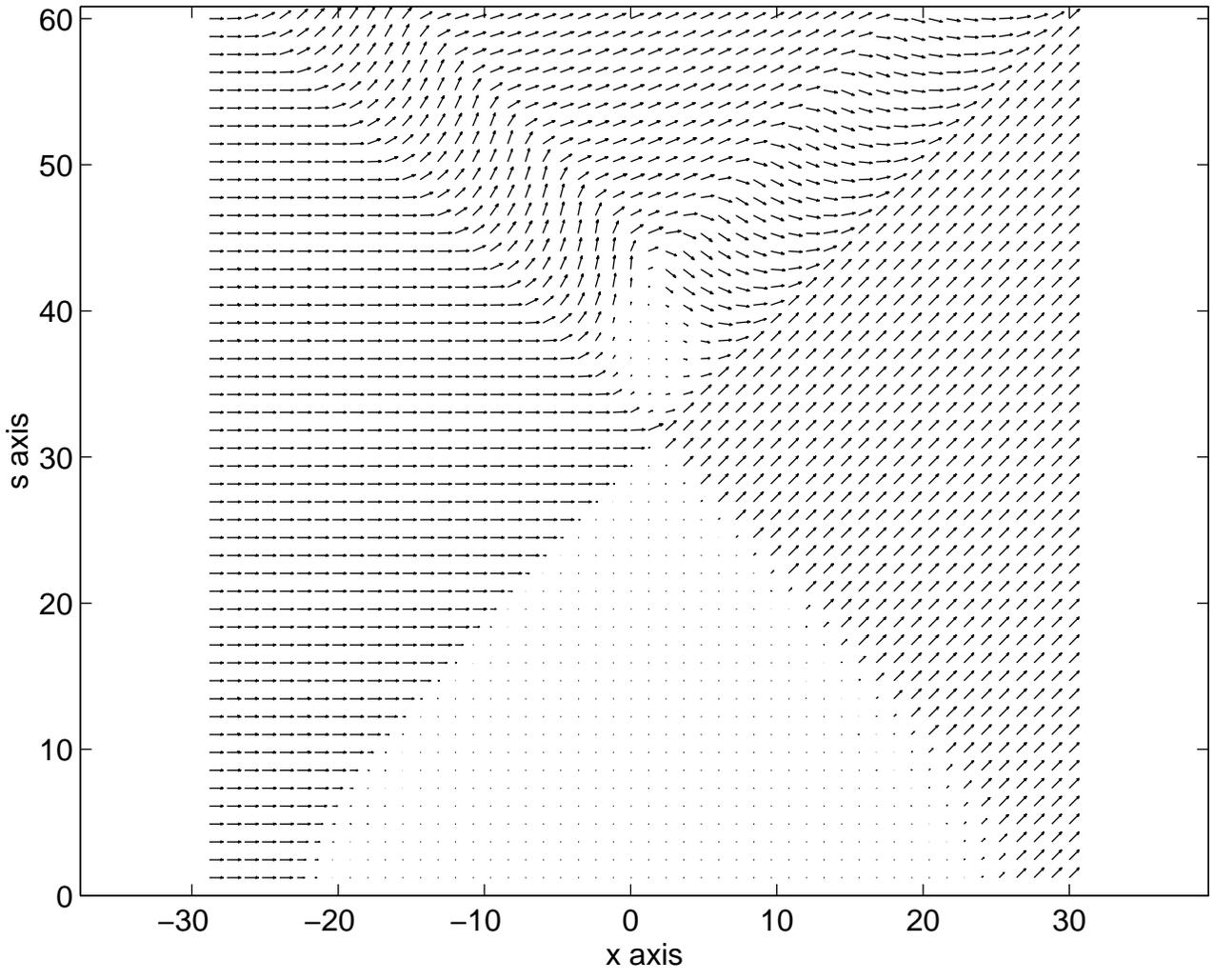
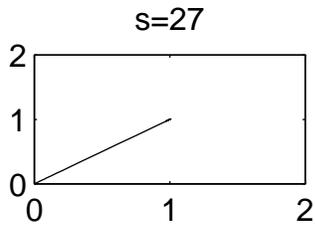
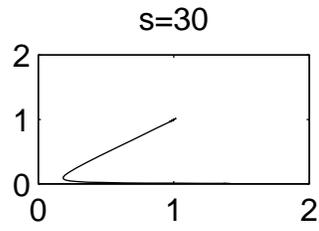
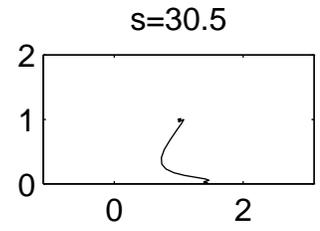
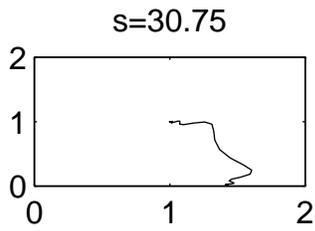
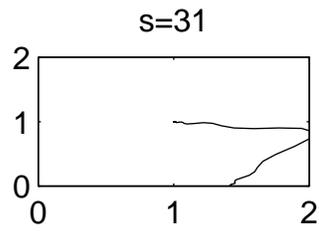
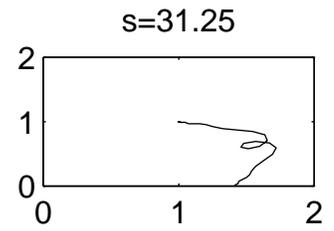
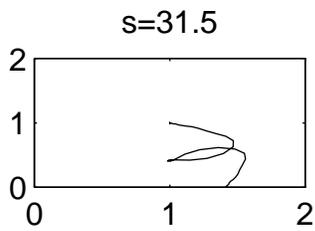
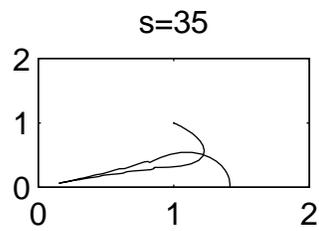
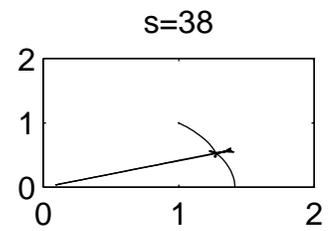
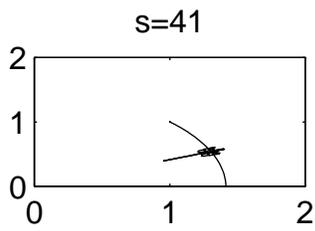
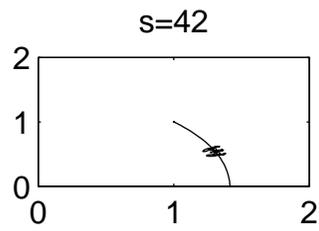
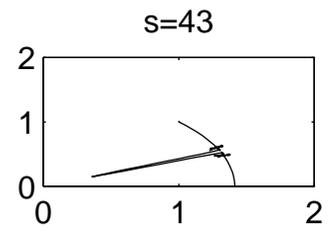
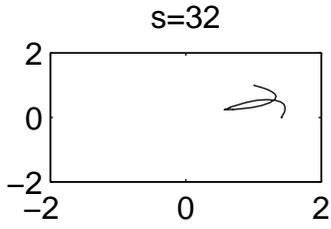
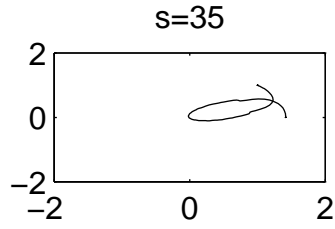
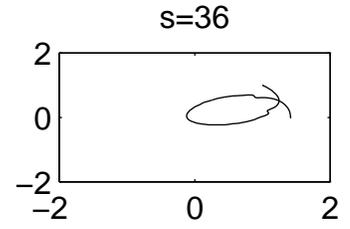
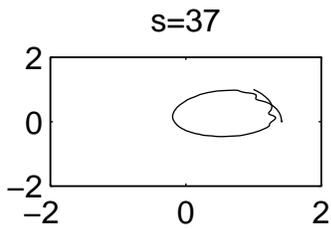
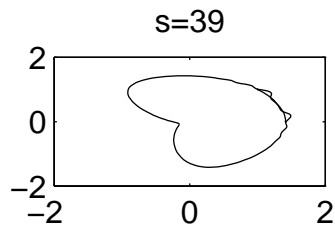
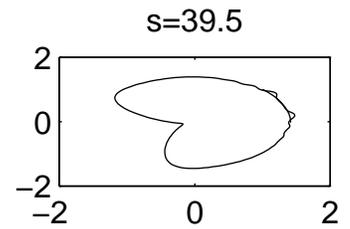
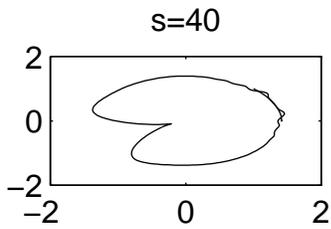
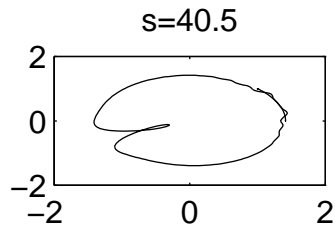
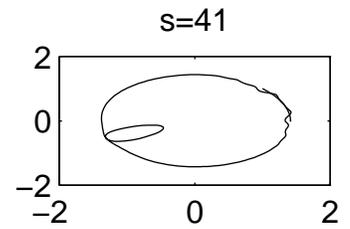
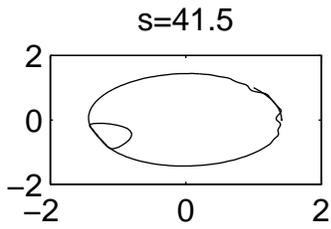
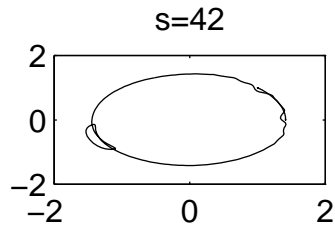
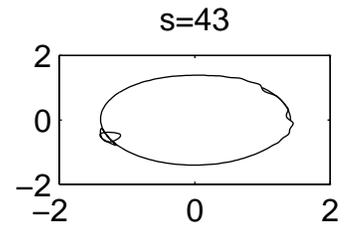
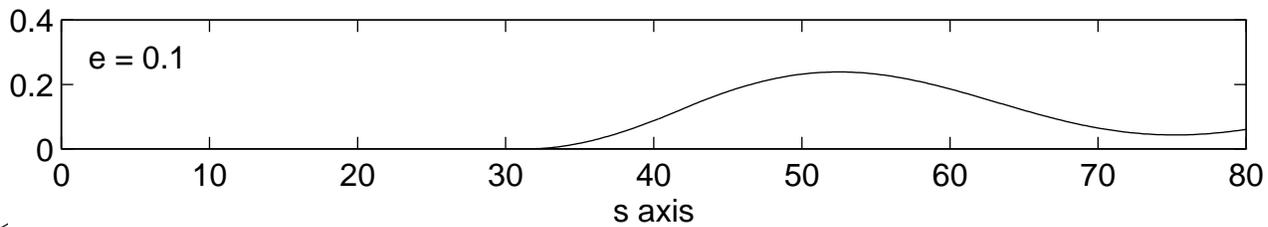
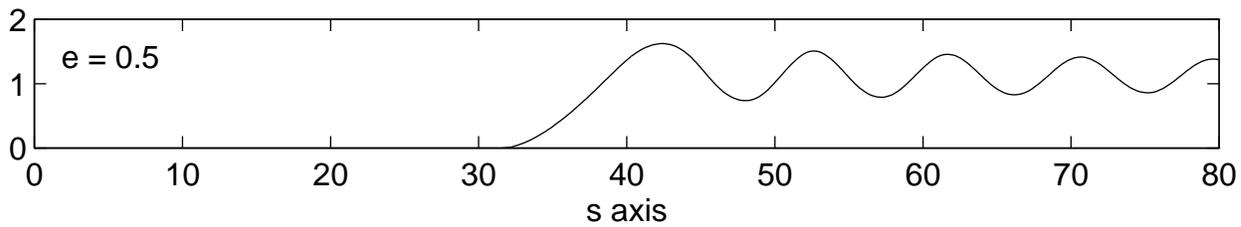
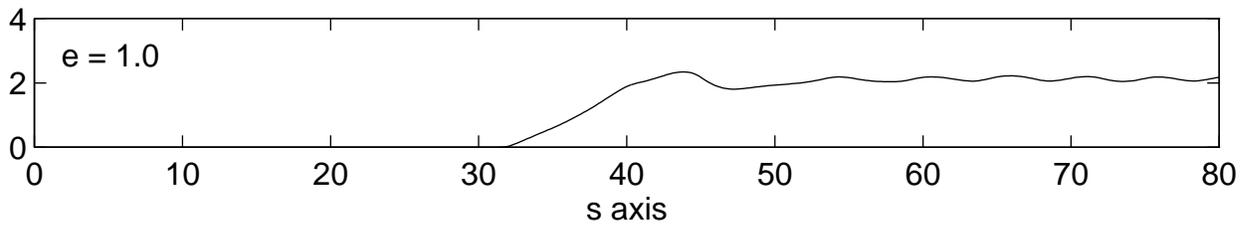
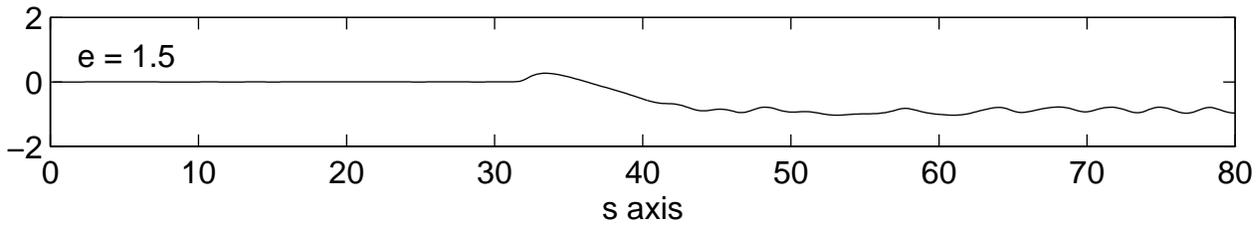
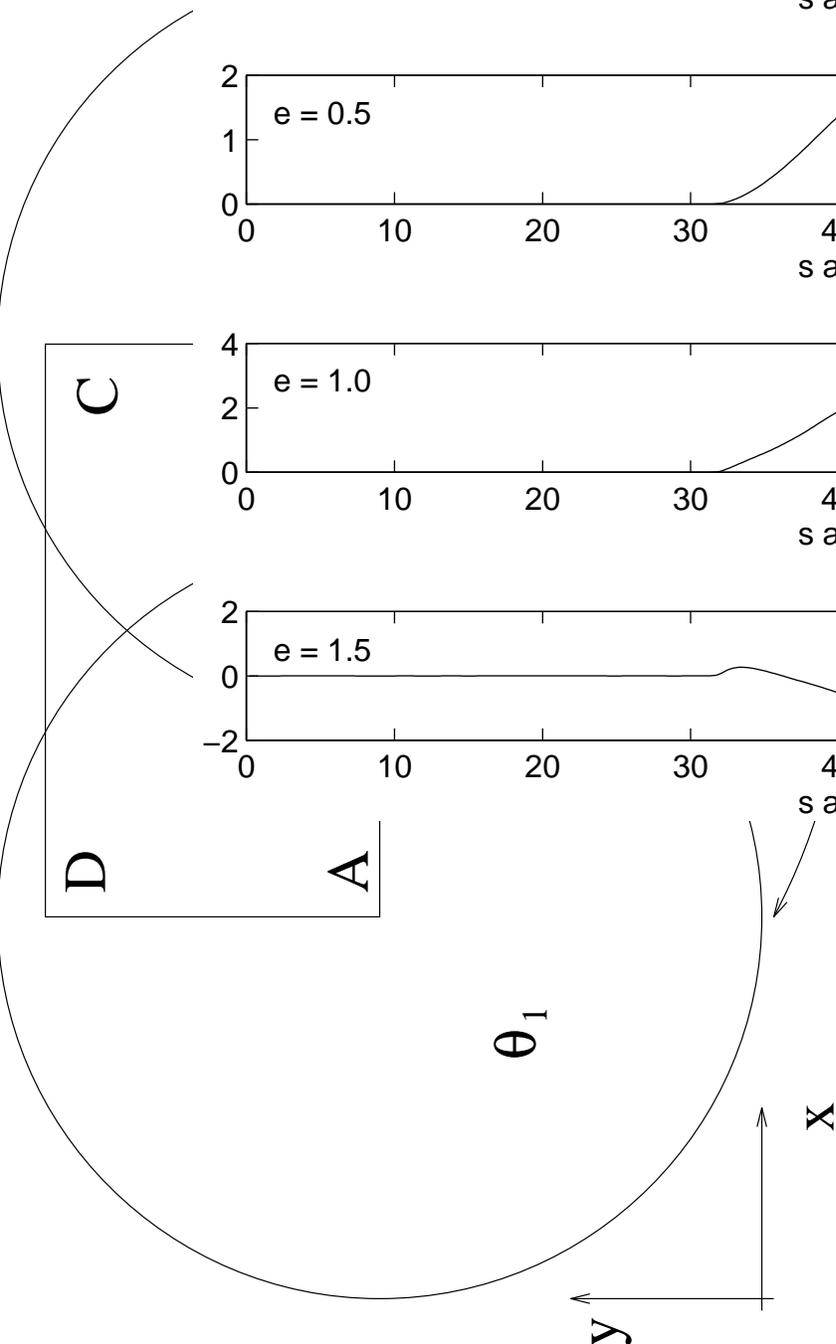
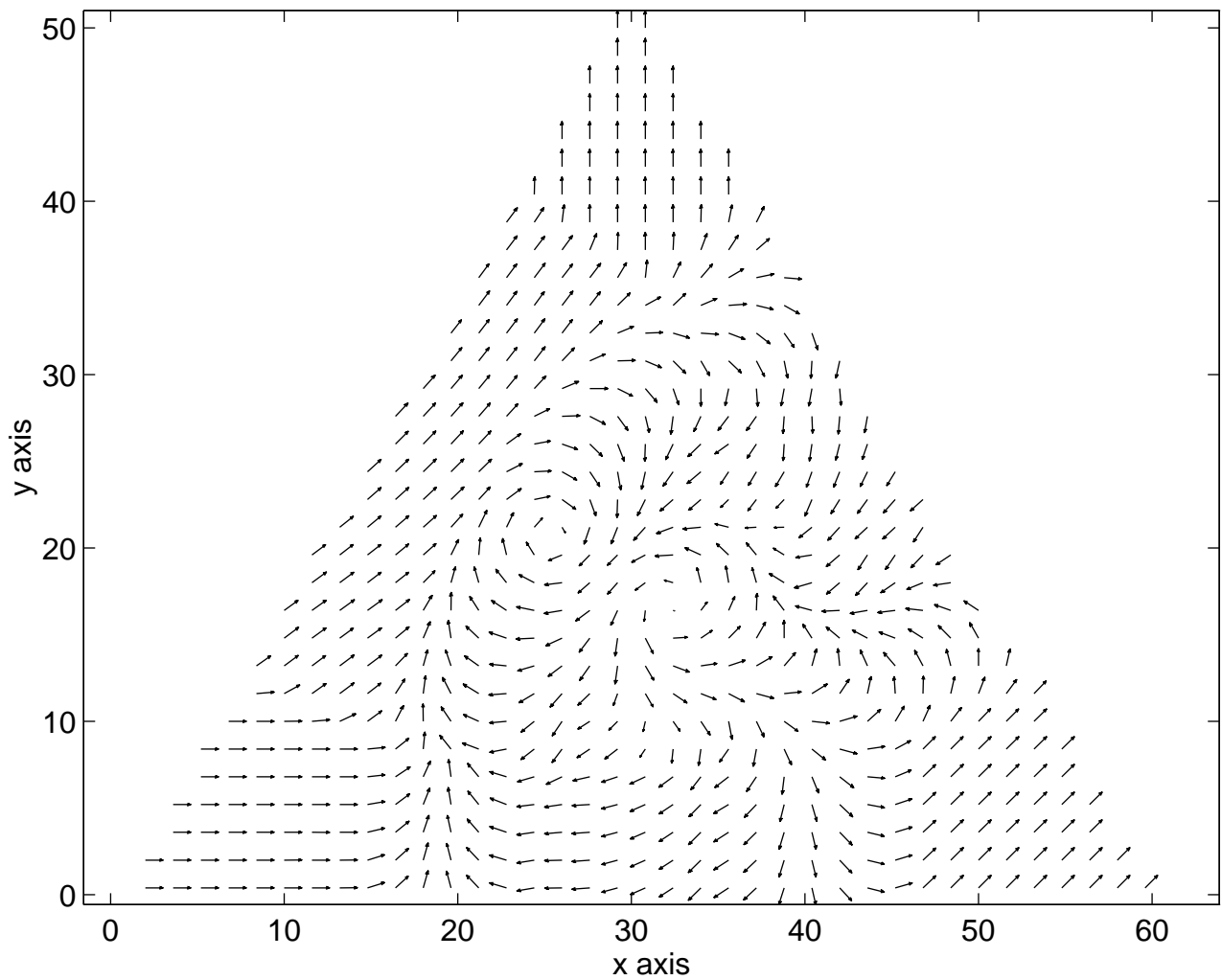
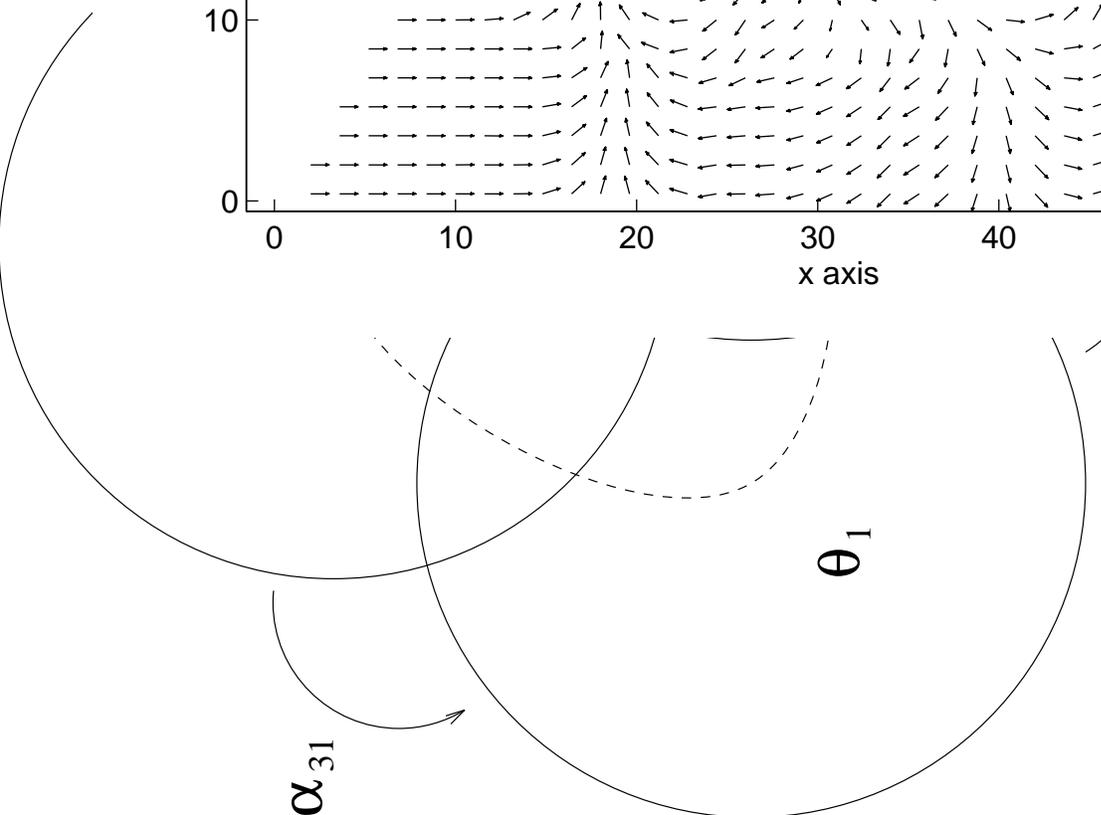
\end{document}